\documentclass[reqno,10pt]{amsart}
\usepackage{xspace}
\newcommand{\bd}{\begin{definition}}                
\newcommand{\ed}{\end{definition}}                  
\newcommand{\bc}{\begin{corollary}}                 
\newcommand{\ec}{\end{corollary}}                   
\newcommand{\bl}{\begin{lemma}}                     
\newcommand{\el}{\end{lemma}}                       
\newcommand{\bp}{\begin{proposition}}            
\newcommand{\ep}{\end{proposition}}                
\newcommand{\bere}{\begin{remark}}                  
\newcommand{\ere}{\end{remark}}                     

\DeclareMathOperator{\de}{d}                        
\newcommand{\R}{\ensuremath{\mathbb{R}}\xspace}     
\newcommand{\inte}{\int_0^1\!\!}
\newcommand{\ptl}{\ensuremath{\omega}\xspace}
\newcommand{\eh}{\ensuremath{F}\xspace}

\newcommand{\dd}{{\rm d}}

\newcommand{\gkk}{\ensuremath{g^{\mathrm{kk}}}}
\newtheorem{theorem}{Theorem}[section]
\newtheorem{corollary}[theorem]{Corollary}
\newtheorem{lemma}[theorem]{Lemma}
\newtheorem{proposition}[theorem]{Proposition}
\theoremstyle{definition}
\newtheorem{definition}[theorem]{Definition}
\theoremstyle{remark}
\newtheorem{remark}[theorem]{Remark}

\begin{document}

\title[Existence of maximizing curves for the charged-particle action]{On the existence of maximizing curves for the charged-particle action}

\author[E. Minguzzi]{Ettore Minguzzi}
\address{Previous address: Dipartimento di Fisica,
Universit\`a di Milano-Bicocca, Piazza della Scienza 3, 20126
Milano, ITALY} \email{minguzzi@mib.infn.it}




\begin{abstract}
The classical Avez-Seifert theorem  is generalized to the case of
the Lorentz force equation for charged test particles with fixed
charge-to-mass ratio. Given two events $x_{0}$ and $x_{1}$, with
$x_{1}$ in the chronological future of $x_{0}$, and a ratio $q/m$,
it is proved that a timelike connecting solution of the Lorentz
force equation exists provided there is no null connecting
geodesic and the spacetime is globally hyperbolic. As a result,
the theorem answers affirmatively to the existence of timelike
connecting solutions for the particular case of Minkowski
spacetime. Moreover, it is proved that there is at least one
$C^{1}$ connecting curve that maximizes the functional
$I[\gamma]=\int _{\gamma} \dd s+q/(mc^2) \omega$ over the set of
$C^{1}$ future-directed non-spacelike connecting curves.
\end{abstract}
\maketitle

\section{Introduction}

Let $\Lambda$ be a  Lorentzian manifold endowed with the metric
$g$ having signature $(+ - - -)$. A  point particle of rest mass
$m$ and electric charge $q$, moving in the electromagnetic field
$F$, has a timelike worldline that satisfies the {\em Lorentz
force equation} (cf. \cite{mtw})
\begin{equation} \label{lorentz}
 D_s \left(\frac{\dd x}{\dd
s}\right)=\frac{q}{mc^2}\hat F(x)\left[\frac{\dd x}{\dd s}\right].
\end{equation}
 Here $x=x(s)$ is the world line of the particle parameterized
with proper time, $\frac{\de x}{\de s}$ is the four-velocity,
$D_s\left(\frac{\de x}{\de s}\right)$ is the covariant derivative
of $\frac{\de x}{\de s}$ along $x(s)$ associated to the
Levi-Civita connection of $g$, and $\hat F(x)[\cdot]$ is the
linear map  on $T_x \Lambda$ defined by
\begin{equation*}
g(x)[v,\hat F(x)[w]] = F(x)[v,w] ,
\end{equation*}
for any $v,\ w\in T_x\Lambda$.

Let $x_0$ and $x_1$ be two chronologically related events, $x_1
\in I^{+}(x_0)$.  If the manifold $\Lambda$ is globally
hyperbolic, the Avez-Seifert theorem \cite{BEE,AV,HS} assures the
existence of at least a timelike connecting solution of the
Lorentz force equation in the $q=0$ case. We are looking for a
suitable generalization to $q/m \ne 0$ cases.

Works in this direction \cite{CMI,CM} have shown that in a
globally hyperbolic manifold $\Lambda$, and for an exact
electromagnetic field $F=d \omega$ (i.e. in absence of monopoles),
 connecting solutions exist   for any ratio $q/m$ in a suitable
neighborhood $[-R,R]$. $R$ is a gauge invariant quantity that
depends on the extremals $x_0$ and $x_1$ and on the potential
one-form. That result was satisfying from the physical point of
view since  for  sufficiently weak field, compatible with the
absence of quantum pair creation effects, the electron's
charge-to-mass ratio  is less than $R$.

From a mathematical point of view, however, the problem in the
strong field case was still open. Here we prove that under the
same conditions as above $R=+\infty$ provided there is no null
connecting geodesic.

Like in previous papers on the subject \cite{CM,CMI}, the strategy
is to introduce a Kaluza-Klein spacetime \cite{L,K} and to regard
the solutions of the Lorentz force equation as projections of null
geodesics of a higher dimensional manifold. In this way  one can
take advantage of causal techniques. Here the reference text for
most notations and results on causal techniques is \cite{HE}.

So assume that $\eh$ is an exact two-form and let \ptl be a
potential one-form for \eh. Let us consider a trivial bundle
$P=\Lambda \times \mathbb{R}$, $\pi : P \to \Lambda$, with the
structure group $T_{1}:$ $ b \in T_{1}$, $p=(x,y)$, $p'=p b=(x,
y+b)$, and $\tilde\ptl$ the connection one-form on $P$:
\[
\tilde\ptl=i(\dd y + \frac{e}{\hbar c} \ptl).
\]
Here  $y$ is a dimensionless coordinate on the fibre, $-e \,
(e>0)$ is the electron charge and $\hbar=h/2\pi$, with $h$ the
Planck constant. Henceforth we will denote by $\bar \ptl$ and
$\bar\eh$, respectively the one-form $\frac{e}{\hbar c} \ptl$ and
the two-form $\frac{e}{\hbar c} \eh$. Let us endow $P$ with the
Kaluza-Klein metric
\begin{equation}
\gkk=g+a^{2}  \tilde\omega^{2}  \label{kk}
\end{equation}
or  equivalently, using the notation $z$ for the points in $P$ and
the identification $z=(x,y)\in \Lambda\times\R$,
\[
\gkk(z)[w,w]=\gkk(x,y)[(v,u),(v,u)]=g(x)[v,v]- a^{2}(u +
\bar\ptl(x)[v])^{2},
\]
for every $w=(v,u)\in T_x \Lambda\times \R$. The positive constant
$a$ has the dimension of a length and  has been introduced for
dimensional consistency of definition \eqref{kk}.

Let $x_1$ be an event in the chronological future of $x_0$. The
set  $\mathcal{N}_{x_0,x_1}$, includes the $C^1$ future-pointing
non-spacelike connecting curves. With {\em connecting curve} we
mean a map $x$ from an interval $[a,b]\subset \R$ to $\Lambda$
such that $x(a)=x_{0}$ and $x(b)=x_{1}$ and any other map $w$ such
that $w=x\circ\lambda$ with $\lambda$ a
 $C^1$ function from an interval
$[c,d]$ to the interval $[a,b]$, having positive derivative.

The functional $I[\gamma]$  defined on the space
$\mathcal{N}_{x_0,x_1}$ is
\[
I[\gamma](x_0,x_1)=\int _{\gamma} (\dd s+\frac{q}{mc^2}  \omega).
\]
The timelike solutions of the Lorentz force equation
(\ref{lorentz}), if they exists, are  critical points of this
functional as it follows from a computation of the Euler-Lagrange
equation.

Let us now consider the geodesics over $P$. They are $C^{1}$
curves $z(\lambda)=(x(\lambda),y(\lambda))$ that are critical
points of the functional
\begin{equation*}
S=S(z)=\inte\frac{1}{2} \gkk(z(\lambda))[\dot z(\lambda),\dot
z(\lambda)]\de \lambda.
\end{equation*}
Taking into account that $\gkk$ is independent of $y$ we find that
the following quantity is conserved
\[
p_{z}=-a^{2}(\dot{y} + \bar\ptl(x)[\dot{x}]).
\]
Moreover taking variations with respect to the variable $x$ we
obtain
\begin{equation} \label{x}
D_{\lambda} \dot x=p_{z}\hat{\bar{F}}(x)[\dot x].
\end{equation}
If $x$ is non-spacelike we define
\[
g(x)[\dot x,\dot x] =C^{2}.
\]
Moreover, since $z$ is a geodesic, $\gkk(z)[\dot z,\dot z]$ is
conserved too and
\begin{equation}\label{gkkconst}
\gkk(z)[\dot z,\dot z]=C^{2}-\frac{p_{z}^{2}}{a^{2}}.
\end{equation}
From this formula it follows that if $z$ is timelike
(non-spacelike) then $x$ is timelike (non-spacelike). If $z$ is a
null geodesic then $C^{2}  = p_{z}^{2}/a^{2}$ and $x$ is timelike
iff $p_z\neq 0$.

In case $x$ is timelike its proper time is given by
\[
\dd s= C \dd \lambda,
\]
and parameterizing with respect to proper time Eq. (\ref{x})
becomes
\[
D_s \left(\frac{\de x}{\de s}\right) =
  \frac{p_z}{C}\hat{\bar F}(x)\left[\frac{\de x}{\de s}\right] =
     \frac{p_z}{C}\frac{e}{\hbar c}\hat F(x)\left[\frac{\de x}{\de s}\right].
\]
This is exactly the Lorentz force equation for a charge-to-mass
ratio
\[
\frac{q}{m}=\frac{p_z}{C}\frac{ec}{\hbar}.
\]
Notice that, a solution of Eq. (\ref{x}) must be timelike ($p_z\ne
0$) in order to represent a charged particle. Only in this case it
can be parameterized with respect to proper time.

Our strategy is to search a future-directed null geodesic in $P$
that projects on a connecting timelike curve on $\Lambda$. To this
end we have to choose the following value for $a$
\begin{equation} \label{aa}
a=\vert \frac{p_z}{C}\vert= \frac{\hbar}{ec} \vert \frac{q}{m}
\vert.
\end{equation}

\section{The theorem}

We state the theorem.

\begin{theorem}\label{main}
Let $(\Lambda,g)$ be a time-oriented Lorentzian manifold. Let \ptl
be a one-form ($C^{2}$) on $\Lambda$  and $\eh=\de \ptl$. Assume
that $(\Lambda,g)$ is a globally hyperbolic manifold. Let $x_{1}$
be an event in the chronological future of $x_{0}$ and $q/m$ any
charge-to-mass ratio. There exists at least one future-directed
non-spacelike $C^{1}$ curve $x(\lambda)$
 connecting $x_0$ and $x_1$ that
maximizes the functional $I[\gamma](x_0,x_1)$ on the space
$\mathcal{N}_{x_0,x_1}$.  If $x$ is timelike, once parameterized
with respect to proper time,  it becomes a solution of the Lorentz
force equation (\ref{lorentz}); if it is null, it is a null
geodesic.
\end{theorem}

We need some lemmas.

\bl The manifold $P=\Lambda \times \mathbb{R}$ endowed with the
metric \eqref{kk} is a time-oriented globally hyperbolic
Lorentzian manifold. \el
\begin{proof}
See \cite{CMI} or \cite{CM}.
\end{proof}

\bere\label{E+} Let $E^{+}(p_{0})=J^{+}(p_{0})-I^{+}(p_{0})$,
$p_{0} \in P$. It is well known (see  \cite[p. 112,184]{HE}) that
if $q \in E^{+}(p_{0})$ there exists a null geodesic connecting
$p_{0}$ and $q$. \ere \bl\label{causallysimple} Any globally
hyperbolic Lorentzian manifold $\Lambda$ is causally simple, i.e.
for every compact subset $K$ of $\Lambda$,
$\dot{J}^{+}(K)=E^{+}(K)$,  where $\dot{J}^{+}(K)$ denotes the
boundary of ${J}^{+}(K)$. \el
\begin{proof}
See \cite[p. 188, 207]{HE}.
\end{proof}

\begin{proof}[Proof of Theorem~\ref{main}]
Let $P$ be the Kaluza-Klein principal bundle constructed in the
introduction having $a$ given by Eq. (\ref{aa}). Given a
parameterized curve $\sigma(\lambda): [0,1] \to \Lambda$ belonging
to $\mathcal{N}_{x_0,x_1}$ define its lifts
$\tilde{\sigma}^{+}(\lambda)$ and $\tilde{\sigma}^{-}(\lambda)$ of
starting point $p_0=(x_0,y_0)$ by requiring
$p_{\tilde{\sigma}^{\pm}}= \pm a \int_{\sigma} \dd s$ and
$\tilde{\sigma}^{\pm}(0)=p_0$. In other words
$\tilde{\sigma}^{\pm}(\lambda)=(\sigma(\lambda),
y^{\pm}(\lambda))$ satisfies the condition
\begin{equation} \label{av}
\dot{y}^{\pm}+\bar{\omega}[\dot{\sigma}]=-\frac{p_{\tilde{\sigma}^{\pm}}}{a^2}.
\end{equation}
$\tilde{\sigma}^{\pm}(\lambda)$ is a null curve that depends on
both  $\sigma$ and   its parameterization. Let
$y^{\pm}_1(\sigma)=\tilde{\sigma}^{\pm}(1)$ and $\Delta
y^{\pm}(\sigma)=y^{\pm}_1(\sigma)-y_0$. Integrating Eq. (\ref{av})
over $\sigma$
\begin{equation*}
p_{\tilde{\sigma}^{\pm}}=-a^{2}(\Delta
y^{\pm}(\sigma)+\int_{\sigma} \bar\omega),
\end{equation*}
or
\begin{equation} \label{ciao}
\Delta y^{\pm}(\sigma)=y_{1}^{\pm}(\sigma)-y_{0}=\mp
\frac{1}{a}(\int_{\sigma} \dd s +\frac{(\pm \vert q/m\vert) }{c^2}
\int_{\sigma} \omega ).
\end{equation}
Notice that the final point $p_1=(x_1, y^{\pm}_{1})$ does not
depend on the specific parameterization of $\sigma$. A
maximization on $\mathcal{N}_{x_0,x_1}$ of the functional $I$
relative to the ratio $+ \vert q/m \vert$, corresponds to  a
minimization of $y_{1}^{+}(\sigma)$. Analogously, a maximization
on $\mathcal{N}_{x_0,x_1}$ of the functional $I$ relative to the
ratio $- \vert q/m \vert$ corresponds to  a maximization of
$y_{1}^{-}$. Let
\begin{eqnarray*}
\hat{s}&=& \sup_{\sigma \in  \mathcal{N}_{x_0,x_1}} y_{1}^{-}(\sigma), \\
\bar{s}&=& \inf_{\sigma \in  \mathcal{N}_{x_0,x_1}}
y_{1}^{+}(\sigma),
\end{eqnarray*}
we show that $\hat{s}>\bar{s}$. Indeed,  for a given $\sigma$ we
have
\begin{equation} \label{ai}
y_{1}^{-}(\sigma)-y_{1}^{+}(\sigma)=\frac{2}{a}\int_{\sigma} \dd s
\ge 0
\end{equation}
thus
\begin{equation*}
\hat{s}-\bar{s} \ge \frac{2}{a}\sup_{\sigma \in \
\mathcal{N}_{x_0,x_1}} \int_{\sigma} \dd s=\frac{2l(x_0,x_1)}{a}
\ge 0,
\end{equation*}
with $l(x_0,x_1)$ the Lorentzian distance function. Moreover, both
$\int_\sigma \dd s$ and $\int_{\sigma} |\bar{\omega}|$ are bounded
on $\mathcal{N}_{x_0,x_1}$ \cite{CMI}, therefore $\hat{s}$ and
$\bar{s}$ are finite.

Let $\eta:[0,1] \to P$ be a non-spacelike future-directed $C^{1}$
curve that starts in $p_0$ and ends in $p_1: \pi(p_1)=x_1$. Let
$p_1=(x_1, y_1)$, and consider the projection $x(\lambda)$ of
$\eta(\lambda)$. Since $\eta$ in a non-spacelike curve
\begin{equation*}
g(\dot{x}, \dot{x})-a^{2}(\dot{y}+\bar{\omega}(\dot{x}))^{2} \ge
0.
\end{equation*}
Taking the square-root and integrating over $x(\lambda)$
\begin{equation*}
|y_1-y_0| \le \frac{1}{a} \int_{x} \dd s +
\int_{x}|\bar{\omega}|<M<+\infty,
\end{equation*}
where $M$ is a suitable positive constant. Hence $y_1$ is finite.
Now we consider the set $W=J^{+}(p_0) \cap \pi^{-1}(x_0)$ and
define
\begin{eqnarray*}
\hat{s}'&=& \sup_{p \in  W} y_{1}(p), \\
\bar{s}'&=& \inf_{p \in  W} y_{1}(p).
\end{eqnarray*}
where $y_1(p)$ is defined through $p=(x_1, y_1)$. Since for any
non-spacelike curve $y_1$ is bounded, $\hat{s}'$ and $\bar{s}'$
are bounded too. Since $P$ is globally hyperbolic $J^{+}(p_0)$ is
closed and the set $W$, being limited and closed, is compact. The
points $\hat{p}_1=(x_1,\hat{s}')$ and $\bar{p}_1=(x_1,\bar{s}')$,
being accumulation points, belong to W. Moreover they can't be
points of the open set $I^{+}(p_0)$. Thus, they belong to
$E^{+}(p_0)$ and therefore (remark \ref{E+}) there are two null
geodesics $\hat{\eta}(\lambda)=(\hat{x}(\lambda),
\hat{y}(\lambda))$, $\bar{\eta}(\lambda)=(\bar{x}(\lambda),
\bar{y}(\lambda))$, that join $p_0$ with $\hat{p}_1$ and
$\bar{p}_1$ respectively. Let $\lambda$ be that affine parameter
that has values 0 and 1 at the endpoints. For a null geodesic
$\eta=(x,y)$ as those under consideration, $p_{\eta}$ is
conserved, hence for a suitable choice of sign
$\eta=\tilde{x}^{\pm}$. Let $p_1=(x_1,y_1)$ be its final point.
For the definition of $\hat{s}$ and $\bar{s}$
\begin{equation*}
 \bar{s} \le y_1 \le \hat{s}.
\end{equation*}
But in the case $\eta=\hat{\eta}$ it is $\hat{s}'\ge \hat{s}$,
otherwise there would be a null curve $\beta$ having final point
$\beta(1)$ strictly above $\hat{p}_1$ on $x_1$'s  fiber. This
would be a contradiction since $\beta(1) \in W$ as $\beta$ is a
null curve. With an analogous reasoning for $\bar{\eta}$ we
conclude that
\begin{eqnarray*}
\hat{s}'&=& \hat{s}, \\
\bar{s}'&=& \bar{s}.
\end{eqnarray*}
We are going to show that the right choice of sign for
$\hat{\eta}$ is $-$, that is
\[
p_{\hat{\eta}}=-a \int_{\hat{x}} \dd s \le 0,
\]
 and $\hat{\eta}=\tilde{\hat{x}}^{-}$.

 Assume that $\hat{\eta}=\tilde{\hat{x}}^{+}$
then, from the definition of $\hat{s}' \,\,(=\hat{s})$
\[
\hat{s}'=\tilde{\hat{x}}^{+}(1) \ge \tilde{\hat{x}}^{-}(1).
\]
 Equation (\ref{ai}) implies that $\tilde{\hat{x}}^{-}(1) \ge
\tilde{\hat{x}}^{+}(1)$ where the equality holds if and only if
$\hat{x}$ is a null curve. From the hypothesis we find that
$\hat{x}$ is a null curve and, since in this case both lifts
coincide with the horizontal lift,
$\tilde{\hat{x}}^{-}(\lambda)=\tilde{\bar{x}}^{+}(\lambda)$. Thus
$-$ is always the right sign whereas $+$ is right if and only if
$\hat{x}$ is a null curve, in which case $\tilde{\hat{x}}^{-} =
\tilde{\hat{x}}^{+}$.

With an analogous reasoning for $\bar{\eta}$ we find
\begin{eqnarray*}
\hat{\eta}(\lambda)&=& \tilde{\hat{x}}^{-}(\lambda),\\
\bar{\eta}(\lambda)&=& \tilde{\bar{x}}^{+}(\lambda).
\end{eqnarray*}
We conclude that the functional $I[\gamma](x_0, x_1)$ is maximized
in $\mathcal{N}_{x_0,x_1}$  by $\hat{x}$ if $q/m <0$ or by
$\bar{x}$ if $q/m >0$. The curve $\hat{x}$, being the projection
of a null geodesic, is a connecting solution of Eq. (\ref{x}),
moreover if timelike it is a connecting solution of the Lorentz
force equation with charge-to-mass ratio $-\vert q/m \vert$. If it
is null, from $\vert p_{\hat{\eta}}\vert =a \int_{\hat{x}} \dd
s=0$ and Eq. (\ref{x}) we conclude that it is a null geodesic. An
analogous conclusion holds for $\bar{x}$.
\end{proof}

In many  cases the spacetime $\Lambda$ has the property that no
two chronologically related events are  joined by a null geodesic.
Minkowski spacetime is the most important example.

 \bc \label{M} Let $(M,\eta)$ be the Minkowski spacetime.  Let $\eh$ be
an electromagnetic tensor field (closed two-form).  Let $x_{1}$ be
an event in the chronological future of $x_{0}$ and $q/m$ a
charge-to-mass ratio, then there exist at least one
future-directed timelike solution to \eqref{lorentz} connecting
$x_0$ and $x_1$. \ec
\begin{proof}
Since $M$ is contractible $\eh$ is exact. Moreover, in Minkowski
spacetime, if $x_{1} \in I^{+}(x_{0})$ there is no null geodesic
connecting $x_{0}$ with $x_{1}$.
\end{proof}

\section{Conclusions}
We have shown that in  Minkowski spacetime the existence of at
least a timelike connecting solution to the Lorentz force equation
is assured by corollary \ref{M}. Notice that theorem \ref{main}
holds more generally for any chronologically related pair $x_0$,
$x_1$, belonging to a globally hyperbolic set $N \subset M$.  Thus
in a generic spacetime  the existence of a timelike connecting
solution to the Lorentz force equation is assured whenever $x_0$
and $x_1$ belong to a globally hyperbolic set and there is no
connecting null geodesic. Finally, we have proved the existence of
at least one $C^{1}$ connecting curve that maximizes the
functional $I[\gamma]=\int _{\gamma} \dd s+q/(mc^2) \omega$ over
the set of $C^{1}$ future-directed non-spacelike connecting
curves.


\end{document}